\documentclass[preprint,showpacs,amsmath,amssymb]{revtex4}
\begin{document}

\title{Dynamic equations for three different qudits in a magnetic
field}


\author{E. A. Ivanchenko}
\email[]{yevgeny@kipt.kharkov.ua}
\affiliation{{National Science Center \textquotedblleft{Kharkov}
Physicotechnical Institute\textquotedblright{}, Institute for
Theoretical Physics,
 \\
 Akademicheskaya Str. 1, Kharkov 61108, Ukraine}}


\date{\today}

\begin{abstract}
A closed system of equations for the local Bloch vectors and spin
correlation functions of three magnetic qudits, which are in an
arbitrary, time-dependent, external magnetic field, is obtained
using decomplexification of the Liouville-von Neumann equation.
The algorithm of the derivation of the  dynamic equations   is
presented.
  In the basis convenient for the important physical
 applications  structure constants of  algebra su(2S+1) are calculated.
\end{abstract}
 \pacs{03.67.-a, 03.67.Ac, 03.67.Mn}

\maketitle
\section{Introduction}
\indent At present it is not known which experiment will lead to
the  reliable, prototypical quantum computing device. Quantum
systems with two states, called qubits, are taken to be the basic
unit for quantum information processing. The quantum systems
realised on coupled qubits are widely investigated for the purpose
of creation of a quantum computer. In the same time there is a
possibility to solve this problem, using the systems consisting
from coupled qudits (multilevel systems). Qudits possess a number
of properties which differ from properties qubits and may have
advantages  for quantum information processing. For example, two
qutrits can be entangled  more strongly than two qubits
\cite{Jing-LingChenDagomirKaszlikowskiKwekZukowskiOh}.
  The main  purpose of this study is the derivation of   dynamic equations
  for  three  qudit system,
which is placed in a magnetic field, for entanglement study
\cite{MingLiShaoMingFeiZhiXiWang}
  and other applications.
 The article is organized as follows. Section II contains the
 model Hamiltonian
including the three-particle interactions.
   In section III the Liouville-von Neumann equation  for the matrix  density of three
   qudits in the  variable
 magnetic field
     is obtained in the Bloch's representation
 in terms of   local vectors  and spin correlation functions taking
 into account three-particle interactions.
  We describe   conservation laws which   control numerical calculations effectively.
   The Appendix describes the Hermitian basis and analytic formulas for structure constants.

 \section{Model Hamiltonian}
\indent The Hamiltonian of three coupled different  qudits
(particles with spin $ 1/2,\,1,\, 3/2, \,2,\,\dots$) in the
external ac magnetic field $\textbf{h}=(h_1,h_2,h_3)$, where
$h_1,h_2,h_3$ are the Cartesian components of the external
magnetic field, we will present in a form of decomposition on a
complete set orthogonal Hermitian matrices $C_\alpha $, operating
in the Hilbert space of three qudits $ H^{S_1} \otimes
H^{S_2}\otimes H^{S_3}$  with $(2S_1+1)(2S_2+1)(2S_3+1)$ dimension
\begin{equation}\label{eq:1}
  \hat{H}=\frac{1}{2}
  h_{\varrho\sigma\tau}C^{S_1}_\varrho\otimes  C^{S_2}_\sigma \otimes C^{S_3}_\tau,
\end{equation}
where $S_1, S_2, S_3$ are qudit spins, $\otimes$ - denotes a
direct product. $C^{S_1}_\varrho, C^{S_2}_\sigma , C^{S_3}_\tau$
are the Hermitian matrices operating in spaces $ H^{S_1}$, $
H^{S_2}$, $H^{S_3}$ accordingly (see Appendix).  The functions
$h_{\varrho\sigma\tau}$ contain one-~,  two- and three-qudit
interactions \cite{Pachos}, \cite{PacosKnight},
\cite{ZhangPengSuter}.

\section{Decomplexificatin of  the Liouville-von Neumann equation}
 The Liouville-von Neumann equation for the density matrix
$\rho$, describing the dynamics of a 3-qudit system, has the form
\begin{equation}\label{eq:2}
  i\partial_t\rho=[\hat{H},\rho], \, \rho(0)=\rho_0, \, \rho^+=\rho, ~{\rm Tr\,}\rho=1.
\end{equation}

 It is convenient to rewrite the equation  Eq.~(\ref{eq:2})\,\,
 having presented the
density matrix $ \rho $ as well as the Hamiltonian $ \hat {H} $,
in the form of decomposition on a complete set orthogonal
Hermitian matrices $C_\alpha$

\begin{equation}\label{eq:3}
  \rho=\frac{1}{ c_1(2S_1+1)c_2(2S_2+1)c_3(2S_3+1)}
  R_{\alpha\beta\gamma}C^{S_1}_\alpha\otimes  C^{S_2}_\beta \otimes C^{S_3}_\gamma,
\end{equation}
where $ c_{1,2,3}= \sqrt{ \frac{S_{1,2,3}(S_{1,2,3}+1)}{3}} $ .
  We define the three coherence Bloch vectors  $R_{m00},R_{0n0},R_{00p}$
, which are widely used in the theory of magnetic
resonance  characterize the local properties of the individual
qudits,
\begin{subequations}\label{eq:4}
\begin{equation}\label{eq:4a}
  c_1c_2c_3 R_{m00}={\rm Tr\,} \rho\,C^{S_1}_m\otimes C^{S_2}_0\otimes C^{S_3}_0,
\end{equation}
\begin{equation}\label{eq:4b}
  c_1c_2c_3 R_{0n0}={\rm Tr\,} \rho\,C^{S_1}_0\otimes C^{S_2}_n\otimes C^{S_3}_0,
\end{equation}
\begin{equation}\label{eq:4c}
  c_1c_2c_3 R_{00p}={\rm Tr\,} \rho\,C^{S_1}_0\otimes C^{S_2}_0\otimes C^{S_3}_p.
\end{equation}
\end{subequations}
 while the tensors $R_{mn0},R_{m0p},R_{0np},R_{mnp}$
\begin{subequations}\label{eq:5}
\begin{equation}\label{eq:5a}
  c_1c_2c_3 R_{mn0}={\rm Tr\,} \rho\,C^{S_1}_m\otimes C^{S_2}_n\otimes C^{S_3}_0 ,
\end{equation}
\begin{equation}\label{eq:5b}
  c_1c_2c_3 R_{m0p}={\rm Tr\,} \rho\,C^{S_1}_m\otimes C^{S_2}_0\otimes C^{S_3}_p,
\end{equation}
\begin{equation}\label{eq:5c}
  c_1c_2c_3 R_{0np}={\rm Tr\,} \rho\,C^{S_1}_0\otimes C^{S_2}_n\otimes C^{S_3}_p,
\end{equation}
\begin{equation}\label{eq:5d}
  c_1c_2c_3 R_{mnp}={\rm Tr\,} \rho\,C^{S_1}_m\otimes C^{S_2}_n\otimes
  C^{S_3}_p,
\end{equation}
\end{subequations}
 describe the spin correlations.
In the formulas  Eq.~(\ref{eq:4}), Eq.~(\ref{eq:5}) $C^{S_i}_0
\equiv C^{S_i}_{0,z}=\sqrt{\frac{S_{i}(S_{i}+1)}{3}}E_{S_i}$,
$E_{S_i}$ is the unit matrix in dimension $(2S_i+1)$  and each of
the Latin indices designates a set of matrices ${
C_{k^{S_i},q^{S_i};x}^{S_i},C_{k^{S_i},q^{S_i};y}^{S_i},C_{k^{S_i};z}^{S_i}
} $ , where $1\leq k^{S_i} \leq 2S_i$, and $1\leq q^{S_i} \leq
k^{S_i}$ in $C_{k^{S_i},q^{S_i};x}^S,C_{k^{S_i},q^{S_i};y}^S$ and
$1\leq k^{S_i} \leq 2S_i$, in $C_{k,z}^S$ in steps of 1, $i=1,2,3$.\\
 \indent Let's formulate the linear algorithm of a decomplexification of  the Liouville-von Neumann equation.\\
1. Insert the Hamiltonian Eq.~(\ref{eq:1}) and the density matrix
Eq.~(\ref{eq:3}), decomposed  on
Hermitian basis into the Liouville-von Neumann equation Eq.~(\ref{eq:2}).\\
2. Multiply  the  Liouville-von Neumann equation by all elements
of the basis $ C^{S_1}_\eta\otimes
C^{S_2}_\theta \otimes C^{S_3}_\vartheta$ in turn.\\
3. Execute operation of a trace taking for each
equation.
 \\
4. Apply the formula  ${\rm Tr\,}(C^{S_1}_\alpha\otimes
C^{S_2}_\beta \otimes C^{S_3}_\gamma)(C^{S_1}_\epsilon\otimes
C^{S_2}_\varepsilon \otimes C^{S_3}_\zeta)( C^{S_1}_\eta\otimes
C^{S_2}_\theta \otimes C^{S_3}_\vartheta) ={\rm
Tr\,}C^{S_1}_\alpha C^{S_1}_\epsilon C^{S_1}_\eta\, {\rm
Tr\,}C^{S_2}_\beta C^{S_2}_\varepsilon C^{S_2}_\theta\,
{\rm Tr\,}C^{S_3}_\gamma C^{S_3}_\zeta C^{S_3}_\vartheta$.\\
5. Express a trace from the three matrices through structure
constants according to formulas
Eqs.~(\ref{eqa:4},\ref{eqa:5}).\\
6. As a result the  use  of the structure constants symmetry the
real terms in each equation  are cancelled out, and the purely
imaginary terms are duplicated.\\
7. The imaginary unit $i$ is  cancelled.\\
This algorithm  is easy to apply for the system of more than 3 qudits.\\
In terms of the functions $R_{\alpha\beta\gamma}$  the
Liouville-von Neumann equation becomes real and comprises a closed
system of  first-order differential equations for the local Bloch
vectors and spin correlation functions

\begin{subequations}\label{eq:6}
\begin{equation}\label{eq:6a}
\partial_t R_{m00}=c_2c_3e^{S_1}_{jim}(h_{j00}R_{i00}+h_{jk0}R_{ik0}+
h_{j0k}R_{i0k} +h_{jkr}R_{ikr}),
\end{equation}
\begin{equation}\label{eq:6b}
\partial_t R_{0n0}=c_1c_3e^{S_2}_{ikn}(h_{0i0}R_{0k0}+h_{ji0}R_{jk0}+
h_{0ij}R_{0kj} +h_{jir}R_{jkr}),
\end{equation}
\begin{equation}\label{eq:6c}
\partial_t R_{00p}=c_1c_2e^{S_3}_{ijp}(h_{00i}R_{00j}+h_{k0i}R_{k0j}+
h_{0ki}R_{0kj}+h_{lki}R_{lkj}),
\end{equation}

\begin{equation}\label{eq:6d}
\begin{split}
\partial_t R_{mn0} &=c_2c_3e^{S_1}_{jim}h_{j00}R_{in0} +c_1c_3e^{S_2}_{jkn}h_{0j0}R_{mk0}+
c_2c_3e^{S_1}_{jim}h_{jn0}R_{i00}+c_1c_3e^{S_2}_{qkn}h_{mq0}R_{0k0}+
\\
& \quad
c_3(e^{S_1}_{jim}g^{S_2}_{kqn}+g^{S_1}_{jim}e^{S_2}_{kqn})h_{jq0}R_{ik0}+
c_2c_3e^{S_1}_{jim}h_{j0q}R_{inq}+c_1c_3e^{S_2}_{jkn}h_{0jq}R_{mkq}+
\\
& \quad
+c_2c_3e^{S_1}_{jim}h_{jnq}R_{i0q}
+c_1c_3e^{S_2}_{ikn}h_{miq}R_{0kq}+ c_3(e^{S_1}_{jim}g^{S_2}_{lkn}
   +g^{S_1}_{jim}e^{S_2}_{lkn})h_{jlq}R_{ikq},
 \end{split}
\end{equation}
\begin{equation}\label{eq:6e}
\begin{split}
\partial_t R_{m0p} &=c_2c_3e^{S_1}_{jim}h_{j00}R_{i0p}+c_1c_2e^{S_3}_{kqp}h_{00k}R_{m0q}+
c_2c_3e^{S_1}_{jim}h_{j0p}R_{i00}+
c_1c_2e^{S_3}_{kqp}h_{m0k}R_{00q}+
\\
& \quad c_2(e^{S_1}_{jim}g^{S_3}_{kqp}+
g^{S_1}_{jim}e^{S_3}_{kqp})h_{j0k}R_{i0q}+c_2c_3e^{S_1}_{jim}h_{jk0}R_{ikp}+
c_1c_2e^{S_3}_{lqp}h_{0kl}R_{mkq}+
\\
& \quad
c_1c_2e^{S_3}_{lqp}h_{mkl}R_{0kq}+c_2c_3e^{S_1}_{jim}h_{jkp}R_{ik0}+
c_2(e^{S_1}_{jim}g^{S_3}_{rqp}+g^{S_1}_{jim}e^{S_3}_{rqp})h_{jkr}R_{ikq},
 \end{split}
\end{equation}
\begin{equation}\label{eq:6f}
\begin{split}
\partial_t R_{0np} &=c_1c_3e^{S_2}_{ikn}h_{0i0}R_{0kp}+c_1c_2e^{S_3}_{iqp}h_{00i}R_{0nq}+
c_1c_3e^{S_2}_{ikn}h_{0ip}R_{0k0}+
c_1c_2e^{S_3}_{kqp}h_{0nk}R_{00q}+
\\
& \quad
c_1(e^{S_2}_{ikn}g^{S_3}_{jqp}+g^{S_2}_{ikn}e^{S_3}_{jqp})h_{0ij}R_{0kq}+
c_1c_3e^{S_2}_{lkn}h_{il0}R_{ikp}+c_1c_2e^{S_3}_{lqp}h_{i0l}R_{inq}+\\
& \quad
c_1c_3e^{S_2}_{qkn}h_{iqp}R_{ik0}+c_1c_2e^{S_3}_{lqp}h_{inl}R_{i0q}+
c_1(e^{S_2}_{lkn}g^{S_3}_{rqp}+g^{S_2}_{lkn}e^{S_3}_{rqp})h_{irl}R_{ikq},
 \end{split}
\end{equation}
\begin{equation}\label{eq:6g}
\begin{split}
\partial_t R_{mnp} &=
c_2c_3e^{S_1}_{jim}h_{jnp}R_{i00}+c_1c_3e^{S_2}_{qkn}h_{mqp}R_{0k0}+c_1c_2e^{S_3}_{kqp}h_{mnk}R_{00q}
+c_2c_3e^{S_1}_{jim}h_{j0p}R_{in0}+
\\
& \quad
c_1c_3e^{S_2}_{jkn}h_{0jp}R_{mk0}+c_3(e^{S_1}_{jim}g^{S_2}_{qkn}+
g^{S_1}_{jim}e^{S_2}_{qkn})h_{jqp}R_{ik0}+
 c_2c_3e^{S_1}_{jim}h_{jn0}R_{i0p}+\\
 & \quad c_1c_2e^{S_3}_{kqp}h_{0nk}R_{m0q}+c_2(e^{S_1}_{jim}g^{S_3}_{lqp}+
 g^{S_1}_{jim}e^{S_3}_{lqp})h_{jnl}R_{i0q}+
   c_1c_3e^{S_2}_{ikn}h_{mi0}R_{0kp}+\\
& \quad
c_1c_2e^{S_3}_{iqp}h_{m0i}R_{0nq}+c_1(e^{S_2}_{ikn}g^{S_3}_{lqp}+
g^{S_2}_{ikn}e^{S_3}_{lqp})h_{mil}R_{0kq}+
c_2c_3e^{S_1}_{jim}h_{j00}R_{inp}+\\
& \quad c_1c_3e^{S_2}_{jkn}h_{0j0}R_{mkp}+
c_3(e^{S_1}_{jim}g^{S_2}_{lkn}+g^{S_1}_{jim}e^{S_2}_{lkn})h_{jl0}R_{ikp}+
c_1c_2e^{S_3}_{jqp}h_{00j}R_{mnq}+\\
& \quad
c_2(e^{S_1}_{jim}g^{S_3}_{lqp}+g^{S_1}_{jim}e^{S_3}_{lqp})h_{j0l}R_{inq}+
c_1(e^{S_2}_{jkn}g^{S_3}_{lqp}+g^{S_2}_{jkn}e^{S_3}_{lqp})h_{0jl}R_{mkq}+
\\
 & \quad
(e^{S_1}_{jim}g^{S_2}_{lkn}g^{S_3}_{rqp}+
g^{S_1}_{jim}e^{S_2}_{lkn}g^{S_3}_{rqp}+
g^{S_1}_{jim}g^{S_2}_{lkn}e^{S_3}_{rqp}-e^{S_1}_{jim}e^{S_2}_{lkn}e^{S_3}_{rqp})h_{jlr}R_{ikq}.
 \end{split}
\end{equation}
\end{subequations}
 As $i\partial_t\rho^n = [\hat {H}, \rho^n] $
$ (n=1,2,3, \dots) $ at unitary evolution there is the numerable
number
 of conservation laws $ {\rm Tr \,} \rho=C_1=1, ~ {\rm
Tr \,} \rho^2=C_2, \dots $, where $C_n  $ are the  constants of
motion, from which only the first $ (2S_1+1) (2S_2+1) (2S_3+1) $
are algebraically independent \cite {Tapia}. From the conservation
of purity, for which $ (\rho^2) _ {ik} \stackrel {\rm def} \equiv
(\rho) _ {ik} $,
  the polynomial (square-law) invariants  are obtained.
The square polynomials also  control  the signs  $R _
{\alpha\beta\gamma} $.
 In the
external dc field the energy of system   is the constant:
 \begin{equation}\label{eq:7}
E={\rm Tr\,}\hat{H}\rho.
\end{equation}
   Unitary evolution preserves
the length of the generalized Bloch vector $ b^{S_1S_2S_3} $
\begin{equation}\label{eq:8}
b^{S_1S_2S_3}
=\sqrt{R_{m00}^2+R_{0n0}^2+R_{00p}^2+R_{mn0}^2+R_{m0p}^2+R_{0np}^2+R_{mnp}^2}.
\end{equation}\\
The qudit   with $ (2S_1+1) $ states in the environment of two
 other qudits  is described by the reduced matrix $ \rho ^ {S_1} $
\begin {equation}\label{eq:9}
\rho ^ {S_1} = \frac {1} {c_1 (2S_1+1)} (R _{000} C ^{S_1}_0+R _
{m00} C ^ {S_1} _m),
\end {equation}
in which $R _ {000} \equiv1$, and functions $R _ {m00} $ are
determinated by the system solution Eq.~(\ref{eq:6}), as the
equations
for the reduced matrices are not closed.\\
 \indent For two different
coupled qudits we have $\hat{H}=\frac{1}{2} h_{\alpha
\beta}C_{\alpha}^{S_1} \otimes C_{\beta}^{S_2}$,
\begin{equation}\label{eq:10}
  \rho=\frac{1}{c_1(2S_1+1)c_2(2S_2+1)
  }R_{\gamma\delta}C_{\gamma}^{S_1}\otimes C_{\delta}^{S_2},\,
  R_{00}=1.
 \end{equation}
 \\
  The dynamic equation Eq.~(\ref{eq:2}) for two different qudits takes on the real
form in terms of the functions $R _ {m0},R _ {0m},R _ {mn} $ as a
closed system of differential equations
Eqs.~(\ref{eq:11},\ref{eq:12},\ref{eq:13}) for the set of initial
conditions:
\begin{equation}\label{eq:11}
\partial_tR_{m0}=c_2e_{pim}^{S_1}(h_{p0}R_{i0}+h_{pl}R_{il}),
\end{equation}
\begin{equation}\label{eq:12}
  \partial_tR_{0m}=c_1e_{pim}^{S_2}(h_{0p}R_{0i}+h_{lp}R_{li}),
  \end{equation}
  \begin{eqnarray}\label{eq:13}
  \partial_tR_{mn} &= e_{pim}^{S_1}\left[c_2(h_{pn}R_{i0}+h_{p0}R_{in})+
  g_{rln}^{S_2}h_{pr}R_{il}\right]+ \nonumber\\
&
 e_{pin}^{S_2}\left[c_1(h_{mp}R_{0i}+
  h_{0p}R_{mi})+
  g_{rlm}^{S_1}h_{rp}R_{li}\right],
  \end{eqnarray}
  where by definition\\
\begin{equation}\label{eq:14}
{\rm Tr\,}\rho C_{\alpha}^{S_1} \otimes
C_{\beta}^{S_2}=c_1c_2R_{\alpha\beta}.
\end{equation}
 The  functions $R_{m0}, R_{0m}$  describe the individual qudits
and the  functions $R_{mn}$ define their correlations.
 The Liouville-von Neumann equation for one
 qudit takes on the real form in terms of the functions $R_{j}
$ as  a closed system of differential equations
\cite{hioe&eberly}:
\begin{equation}\label{eq:15}
\partial_tR_{l}=e_{ijl}^{S_1} h_iR_{j}.
\end{equation}
 The set of equations for 3  qubits  has been obtained in \cite{Ivanchenko}.\\
 \indent
 The set of equations Eq.~(\ref{eq:6})  with the initial  conditions  has wide applications, since
 the magnetic field enters in the form of  arbitrary functions. It allows to make
  numerical calculations  for continuous (a paramagnetic resonance in a continuous mode), as well as
   for pulse modes (a nuclear magnetic resonance).
       By means of this system it is possible to investigate the entanglement dynamics
     of qudits in a magnetic field
   as the entanglement measures are expressed in terms
   of the reduced density matrices or  of populations.
Another important application  of the system Eq.~(\ref{eq:6})  is
quantum approach to  the Carnot cycle \cite{Scovill},
\cite{Carnot}, \cite{Feldmann Kosloff}, \cite{Tova Feldmann and
Ronnie Kosloff}, \cite{RezekKosloff}, when a working body is a
finite spin chain.
\section{Conclusion}
The simple algorithm of the  derivation of  equation  system for
coupled qudits, which are in an arbitrary, time-dependent external
magnetic field  has been presented. It is not necessary for the
basis to be Hermitian since the results of calculations are
independent of the choice of base, but there is the main advantage
with the Hermitian basis. It is that the Liouville-von Neuman
equation not involve  any complex numbers and can be solved using
real algebra. This is not true for non-Hermitian bases. Real
algebra makes numerical calculations faster and simplifies the
interpretation  of the equation system  Eqs.~(\ref{eq:6}).\\
This basis forms a natural basis for calculations on coupled spin
systems  \cite{allard&hard} because all the single-spin operators
are part of the complete basis when the unit operator is part of
the single-spin basis.\\
\begin {acknowledgments}
The author is grateful to Zippa A. A. for  constant invaluable
support.
\end {acknowledgments}
\appendix
\section{}

Let $ \{ C_1^S,C_2^S,...,C_n^S\}$  be a base of su(2S+1) algebra,
where $S=1/2,1,3/2,... $ is the spin quantum number,
$n=(2S+1)^2-1$. We have according to \cite{kimura&kossakowski}
\begin{equation}\label{eqa:1}
C_i^SC_j^S=\frac{c}{d}E\delta_{ij}+z_{ijk}^SC_k^S,~{\rm
Tr\,}C_i^S=0,~ {\rm Tr \,}C_i^SC_j^S=c \delta_{ij},
\end{equation}
\begin{equation}\label{eqa:2}
z_{ijk}^S=g_{ijk}^S+ie_{ijk}^S,
\end{equation}
hence
\begin{equation}\label{eqa:3}
-i[C_i^S,C_j^S]=2e_{ijk}^SC_k^S,
~\{C_i^S,C_j^S\}=\frac{c}{d}E\delta_{ij}+2g_{ijk}^SC_k^S,
\end{equation}
\begin{equation}\label{eqa:4}
e_{ijk}^S=\frac{1}{2ic}{\rm Tr\,}[C_i^S,C_j^S]C_k^S,
\end{equation}
\begin{equation}\label{eqa:5}
g_{ijk}^S=\frac{1}{2c}{\rm Tr\,}\{C_i^S,C_j^S\}C_k^S,
\end{equation} where $d=2S+1$, $E$ is the unit
matrix in  dimension $d \times d$, $c$ is a constant, {\rm Tr\,} is a symbol for
trace. It is easy to see that the structure constants $e_{ijk}^S$ and $g_{ijk}^S$ are
completely antisymmetric  and symmetric in the displacement of any pair of indices.
\subsection{Hermitian basis}
The  structure constants of $su(2S+1)$ algebra have important
physical applications. In order to calculate the structure
constants we have to choose the basis.  The basis is based on
linear combinations of irreducible tensor operators. The matrix
representations of irreducible tensor operators $T_{k,q}^S$
\cite{varsalolovic&moskalev&chersonskij} can be calculated using
the Wigner $3jm$ symbols:
\begin{equation}\label{eqa:6}
T_{k,q}^S=\sqrt{(2S+1)(2k+1)}\sum^S_{m,m'=-S}(-1)^{S-m} \left(
^{\,\,S\,\,\,\,\,\,k\,\,\,S}_{-m\,\,q\,\,\,m'} \right) |S,m><S,m'|,
\end{equation}
where $0\leq k \leq 2S$, and $-k\leq q \leq k$ in steps of 1. The normalization is
such that $T_{0,0}^S=E$. It is known that the Cartesian product operators $S_x,S_y$,
and $S_z$ for  spin $S=\frac{1}{2}$ are Hermitian and can be calculated from
irreducible tensor operators  \cite{ernst&bodenhausen&wokaun}
\begin{subequations}\label{eqa:7}
\begin{equation}\label{eqa:7a}
S_x^{\frac{1}{2}}=\frac{1}{2\sqrt{2}}(T_{1,-1}^{\frac{1}{2}}-T_{1,1}^{\frac{1}{2}})=
\frac{1}{2}\left(\!\!\!%
\begin{array}{cc}
  0 & 1 \\
  1 & 0 \\
\end{array}%
\!\!\!\right),
\end{equation}
\begin{equation}\label{eqa:7b}
S_y^{\frac{1}{2}}=\frac{i}{2\sqrt{2}}(T_{1,-1}^{\frac{1}{2}}+T_{1,1}^{\frac{1}{2}})=
\frac{1}{2}\left(\!\!\!%
\begin{array}{cc}
  0 & -i \\
  i & 0 \\
\end{array}%
\!\!\!\right),
\end{equation}
\begin{equation}\label{eqa:7c}
S_z^{\frac{1}{2}}=\frac{1}{2}T_{1,0}^{\frac{1}{2}}=\frac{1}{2}\left(\!\!\!%
\begin{array}{cc}
  1 & 0 \\
  0 & -1 \\
\end{array}%
\!\!\!\right).
\end{equation}
\end{subequations}
\\
 Allard and H\"{a}rd
\cite{allard&hard} have formed linear combinations of the
irreducible tensor operators not only for single-quantum
coherences, but also for all coherences according to
\begin{subequations}\label{eqa:8}
\begin{equation}\label{eqa:8a}
 C_{k,qx}^S =\sqrt{\frac{S(S+1)}{6}}(T_{k,-q}^S+(-1)^qT_{k,q}^S),~q\neq0,
\end{equation}
\begin{equation}\label{eqa:8b}
C_{k,qy}^S=i\sqrt{\frac{S(S+1)}{6}}(T_{k,-q}^S-(-1)^qT_{k,q}^S),~q\neq0,
\end{equation}
\begin{equation}\label{eqa:8c}
C_{k,z}^S=\sqrt{\frac{S(S+1)}{3}}T_{k,0}^S,~q=0,k\geq1,
\end{equation}
\end{subequations}
\begin{equation}\label{eqa:9}
C_{0,z}^S=\sqrt{\frac{S(S+1)}{3}}E,
\end{equation}
where $1\leq k \leq 2S$, and $1\leq q \leq k$  in $C_{k,qx}^S,C_{k,qy}^S$ and $1\leq
k \leq 2S$,
 in $C_{k,z}^S$ in steps of 1. The matrices Eqs.~(\ref{eqa:8})
 are traceless and their number is equal to
  $(2S+1)^2-1$.  Using
 the well-known relations for the irreducible tensor operators
\begin{equation}\label{eqa:10}
(T_{k,q}^S)^+=(-1)^qT_{k,-q},
\end{equation}
 we can see that  matrices Eqs.~(\ref{eqa:8}) are Hermitian.
Using the formula from \cite{varsalolovic&moskalev&chersonskij}
\begin{equation}\label{eqa:11}
 {\rm Tr\,} T_{k,q}^ST_{k',q'}^S=(-1)^q(2S+1)\delta_{k,k'}\delta_{q,-q'}
\end{equation}
it is easy to show that the basis is normalized so that
$S_x=C_{1,x}^S$, $S_y=C_{1,y}^S$, $S_z=C_{1,z}^S$,
\textit{irrespective of the spin quantum number} $S$, i.e.
\begin{equation}\label{eqa:12}
(C_r,C_s)={\rm Tr\,}C_rC_s=\delta_{r,s}\frac{S(S+1)(2S+1)}{3}.
\end{equation}
The set Eqs.~(\ref{eqa:8},\ref{eqa:9}) is complete. The matrices
$C_{k,z}$ are diagonal
\begin{equation}\label{eqa:13}
[C_{k,z}^S,C_{k',z}^S]=0.
\end{equation}
There also exist the other useful bases \cite{bertmann&krammer},
\cite{Kibler}. From the physical point of view, for important
physical applications the basis \cite{allard&hard} is preferred.
\subsection{Analytic formulas for structure constants}
 There  are
27 combinations  in threes including the repetitions: $XX'X''$,
$XX'Y''$, $XX'Z''$\dots, where $X=C_{k,qx}^S$,  $X'=C_{k',q'x}^S$,
$Y''=C_{k'',q''y}^S$, $Z''=C_{k'',z}^S$ and so on. The use of the
symmetrical properties of the Wigner $3jm$ symbols and the formula
$\textbf{2}.\textbf{4} \,(23) \,
$\cite{varsalolovic&moskalev&chersonskij}
\begin{eqnarray}\label{eqa:14}
{\rm Tr\,}T_{k,q}^S T_{k',q'}^ST_{k'',q''}^S=
(-1)^{2S+k+k'+k''}(2S+1)^{\frac{3}{2}}\nonumber\\
\qquad {}[(2k+1)(2k'+1)(2k''+1)]^{\frac{1}{2}} \{
^{k\,\,\,k'\,\,\,k''}_{S\;\,S\;\;S}\} \left(^{k\;\,k'\;k''}_{q\;\,q'\;q''} \right),
\end{eqnarray}
where $\{ ^{k\,\,\,k'\,\,\,k''}_{S\;\,S\;\;S} \}$ is the $6j$
symbol, allows us after substitution of Eqs.~(\ref{eqa:8}) in
Eq.~(\ref{eq:4}),\,Eq.~(\ref{eq:5}) , to calculate all structure
constants of $su(2S+1)$ algebra. Let us introduce the function
\begin{equation}\label{eqa:15}
F(k,k',k'',S)=\frac{(-1)^{2S}}{\sqrt3}\sqrt{S(S+1)(2S+1) (2k+1)(2k'+1)(2k''+1)}\,\{
^{k\,\,\,k'\,\,\,k''}_{S\;\,S\;\;S} \}.
\end{equation}

All antisymmetric structure constants are zero for $K=k+k'+k''$
even and nonvanishing
 antisymmetric structure constants in terms of $3jm$ and $6j$ symbols have the
explicit form  are presented by formulas
Eqs.~(\ref{eqa:16a},\ref{eqa:16b},\ref{eqa:16c}) for $K$ odd:
\begin{subequations}\label{eq:16}
\begin{equation}\label{eqa:16a}
e_{XX'Y''}^S = -\frac{F}{\sqrt{2}}\!\!\left[(-\!1)^q
\left(^{k\,\,\,k'\,\,\,k''}_{q-q'-q''}\right)+
(-\!1)^{q'}\left(^{k\;\;\;\,k'\;\;\,k''}_{\!-q\,\,\,q'-q''}\right)+(-\!1)^{q''}
\left(^{k\,\,\,k'\,\,\,k''}_{q\;\;q'-q''}\right) \right],
\end{equation}

\begin{equation}\label{eqa:16b}
e_{YY'Y''}^S =
\frac{F}{\sqrt{2}}\!\!\left[(-\!1)^q\left(^{k\;\;\;\,k'\,\,\,k''}_{\!-q\;\,q'\;q''}
\right)+(-\!1)^{q'}\left(^{k\,\,\,\,k'\,\,\,k''}_{q\,-q'\;q''}\right)+
(-\!1)^{q''}\left(^{k\,\,\,\,k'\,\,\,k''}_{q\;\;q'-q''} \right)\right],
\end{equation}
\begin{equation}\label{eqa:16c}
 e_{XY'Z''}^S =
 -F(-\!1)^q\left(^{k\,\,\,\,k'\,\,\,k''}_{q\,\,-q'\,\,0}\right).
\end{equation}
\end{subequations}
All symmetric structure constants are zero for $K=k+k'+k''$ odd
and nonvanishing
 symmetric structure constants in terms of $3jm$ and $6j$ symbols have the
explicit form  are presented by formulas
Eqs.~(\ref{eqa:17a},\ref{eqa:17b},\ref{eqa:17c}) for $K$ even:
\begin{subequations}\label{eqa:17}
\begin{equation}\label{eqa:17a}
g_{XX'X''}^S =
\frac{F}{\sqrt{2}}\!\!\left[(-\!1)^q\left(^{k\;\;\;\,k'\,\,\,k''}_{q\,\,-q'\,-q''}
\right)+(-\!1)^{q'}\left(^{k\;\;\;\,k'\,\,\,k''}_{\!-q\,\,q'\,-q''}\right)+
(-\!1)^{q''}\left(^{k\,\,\,\,k'\,\,\,k''}_{q\;\;q'-q''} \right)\right],
\end{equation}
\begin{equation}\label{eqa:17b}
g_{XY'Y''}^S =
\frac{F}{\sqrt{2}}\!\!\left[-(-\!1)^q\left(^{k\,\,\,\,k'\;\;\,k''}_{q\,\,-q'\,-q''}
\right)+(-\!1)^{q'}\left(^{k\;\,\;k'\;\;\,k''}_{\!-q\,\,q'\,-q''}\right)+
(-\!1)^{q''}\left(^{k\,\,\,\,\,\,k'\,\,\,\,k''}_{\!-q\,-q'\;q''} \right)\right],
\end{equation}
\begin{equation}\label{eqa:17c}
 g_{XX'Z''}^S =g_{YY'Z''}^S= F(-\!1)^q
 \left(^{k\;\;\;k'\,\,\,k''}_{q\,\,-q'\;0}\right),~
 g_{ZZ'Z''}^S = F(-\!1)^q\left(^{k\;\,k'\,k''}_{0\,\,\,0\,\,\,0} \right).
\end{equation}
\end{subequations}
 We have in $X,Y$  \,$1\leq k,k',k'' \leq 2S$,
$1\leq q \leq k, 1\leq q' \leq k',
1\leq q'' \leq k''$ and in $Z$  \, $1\leq k,k',k'' \leq 2S$ in steps of 1. \\
The straightforward calculation confirms that the structure
constants $e_{ijk}^S$ and $g_{ijk}^S$ are completely antisymmetric
and symmetric in the displacement of any pair of operators. In
other words it is $e_{XX'Y''}^S=-e_{XY''X'}^S$, $g_{XX'Z''}^S
=g_{YY'Z''}^S$ and so on.
\end{document}